# A Statistical Model of Pressure Drop Increase with Deposition in Granular Filters

Zhaohui Qin [1] and Richard H. Pletcher [2]

1. Department of Engineering and Computer Science, Cedarville University, Cedarville, OH 45314
   e-mail: gqin@cedarville.edu
2. Department of Mechanical Engineering, Iowa State University, Ames, IA 50011
   e-mail: pletcher@iastate.edu

**Abstract**

As deposits accumulate in a granular filter, pressure drop across the filter bed required to maintain a constant fluid flow rate may increase. Two pressure drop increase patterns had been observed. In slow sand filters pressure drop remains unchanged for a certain period of time then increases exponentially with the volume of filtrate; in granular aerosol filters pressure drop increases linearly with the amount of deposits from the beginning of the filtration process. New concepts of homogeneous and heterogeneous depositions were introduced in this paper. A statistical model based on these new concepts was developed. This non-linear model was able to reproduce both observed pressure drop increase patterns, including the linear one. Excellent agreements between the present model and experimental measurements were obtained. It was concluded that the two pressure drop increase patterns were indeed caused by different deposit distributions rather than different pressure drop increase mechanisms.

**Keywords**: pressure drop, deposition, granular filter, statistical model

## 1. Introduction

Granular filters use granular material to separate small particles from fluids and have many important applications including drinking water purification, waste water treatment, flue gas cleaning, molten metal refinement, radioactive particle removal etc. The increase in pressure drop (head loss) required to maintain a constant flow rate through a granular filter due to particle deposition has been the subject of experimental and theoretical investigations. In this paper major experimental observations and theoretical methodologies are reviewed; two new concepts, namely the homogenous and heterogeneous depositions are proposed; a statistical model based on these concepts is developed and verified.

Experiments revealed two drastically different pressure drop increase patterns in granular filters. In slow sand filters, which are commonly used for water purification, the pressure drop usually remains a constant for a considerable duration as deposits accumulate, then rises exponentially with the volume of filtrate [1][2][3][4]. On the contrary in granular aerosol filters the pressure drop typically increases linearly with the amount of deposit from the very beginning of the filtration process [5][6]. In addition, experiments demonstrated that in slow sand filters the pressure drop increase is concentrated to a thin layer (Schmutzdecke) at the top of the filter bed where the raw water flows into the bed, below which the filter medium remains hydraulically clean, i.e. although deposits present, they affect the pressure drop only marginally and the pressure drop in this region stays essentially the same after even a few years of operation [7]. In contrast, the pressure drop at every depth of a granular aerosol filter increases simultaneously as filtration proceeds [5].

Tremendous efforts have been made to relate these experimental observations to fundamental filtration mechanisms. Because of particle deposition, filter medium structure changes continuously. As most authorities agree [8], such changes include (1) decrease in filter medium porosity and increase in effective filter grain diameter; (2) change in filter grain surface morphology due to non-uniformity of particle deposition and formation of dendrites; (3) clogging part of pores in the granular medium. All these effects contribute to pressure drop increase across the filter.

Effect (1) can be readily evaluated by using Ergun equation [9]. Prediction thus given has been found grossly underestimate the pressure drop increase [10], which is not surprising because deposits only have negligible contribution to pressure drop as long as pores in the granular medium are not clogged [7]. Despite its inability to explain the observed pressure drop increase, Ergun equation has been adopted as the starting point by various investigators [11][12][13] to develop their empirical correlations between pressure drop increase and deposition. These correlations differ a lot in mathematical forms as they were derived from different experimental data. For example the correlation of Mints [11] is linear; while that of Toms [4] is exponential and the one given by Ives [12] is a product of two power-laws.

Trajectory analysis was used extensively by Tien and co-workers [14][15][16][17] to study effect (2) on granular filter performance. By tracking positions of particles randomly released from the inlet of a representative geometry (a constrained tube, for example) which characterizes the "mean" geometrical features of the filter medium, they were able to calculate the deposition location of each deposited particle and subsequently they were able to determine the pressure drop increase across this model space. Their results showed power-law increases in pressure drop with deposits during granular filtration [15][16]. The difference between their prediction and experimental observations might be attributed to clogging, that is effect (3).

Fan *et al.* [18] modeled clogging as a stochastic birth-death process. They treated the granular medium as a large number of interconnected pores and assumed that at every moment during filtration an open pore always has a chance to be blocked, and the probability is proportional to the number fraction of the open pores; at the same time a blocked pore always has a chance to be reopened with a possibility proportional to the number fraction of the blocked pores. The model was able to represent the pressure drop history of a waste water granular filter. However, the two possibility proportionality constants in this model can only be determined by fitting experimental data rather than being related to relevant physical variables.

Based on the existing information, it seems safe to conclude that clogging is the most important reason for the filter pressure drop increase. Compared with clogging the contribution of deposition to the filter pressure drop increase is but minimal. The random nature of the many factors affecting particle deposition implies a statistical treatment of the subject. An important link that connects the mathematical abstraction and physical reality is two new concepts, namely the homogeneous and heterogeneous depositions, which are discussed in the next section.

**2. Homogeneous and Heterogeneous Depositions**

Consider a packed bed of clean granular material through which fluid flows at a constant volume flow rate driven by a pressure gradient. The granular medium can be viewed as layers of pores interconnected in series, as Fig. 1 depicts. The layers are perpendicular to the flow direction. The number density of pores is typically very large.

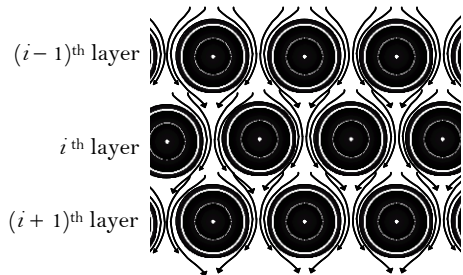

Fig. 1. Granular medium represented as layers of pores

For example for a bed of granules of diameter $d_g = 2\ mm$ and bed porosity $\varepsilon = 0.36$, there are about $1.4 \times 10^4$ pores in $1\ in^3$ of filter volume [19]. As fluid flows through the bed, small particles carried by the fluid may deposit on the surface of pores and on the already deposited particles, or may be absorbed by the organisms in the bed. The rate of particle deposition is affected by a variety of factors such as particle concentration, flow speed, filter grain size, grain surface charge, grain surface morphology, organism concentration in the medium etc. These factors differ from one pore to another. The extent of deposition in different pores of a given layer is therefore different. If the heavier deposited pores have less particle collection ability compared with the lighter deposited pores, the deposition difference among the pores will decrease and a homogeneous, or uniform, deposition distribution will form over this particular layer of pores; if on the contrary the heavier deposited pores have even greater particle collection ability than the lighter deposited pores, the difference in deposition level among pores will increase and a heterogeneous, or non-uniform, deposition distribution will form over this specific layer of pores. Based on logic, a filtration process should always fall into one of these two situations. The details of the physical and/or biological mechanisms leading to these two deposition distribution regimes are indeed irrelevant to the current study (one possible mechanism is given in Appendix A). Instead we are more interested in the inferences of these two deposition distributions.

(1) As fluid-particle suspension flows through a granular filter operating in the homogeneous deposition regime, consecutive uniformly deposited layers form until most particles in the fluid are filtered. Then we should find layers of clean pores. As a consequence, the pressure drop increase as well as deposition in such filters should concentrate to such uniformly deposited layers rather than the whole filter bed. On the other hand, for a granular filter operating in the heterogeneous deposition regime, even as part of pores being clogged at a certain layer, many pores of this layer are still open and of low particle collection ability due to the non-uniformity of the deposition distribution. As a result, much of the suspension can penetrate this layer through such open pores and produce similar partially-clogged deposition patterns in numerous successive layers, even across the whole filter. Therefore we should expect the pressure drop across all such partially-clogged layers, even across the whole filter, to increase simultaneously with time.

Immediately one recognizes the slow sand filters should operate in the homogenous deposition regime and granular aerosol filters typically run in the heterogeneous deposition regime if the present theory is at all reasonable.

(2) In the homogeneous deposition regime the amount of deposits in a clogged pore increases slowly compared with an open pore at the same layer because the heavier deposited pores have less particle collection ability than the lighter deposited pores in this regime. On the other hand, the clogged pores may still actively collect particles if the filter is operating in heterogeneous deposition regime since the heavier deposited pores have higher particle collection ability than the lighter deposited pores. One should notice in the current study "clogged" does not mean "no flow", instead it only means compared with open pores, the clogged pores have significantly less flow under the same pressure gradient.

(3) If we consider the deposit distribution among pores that are actively collecting particles at a specific layer, the homogeneous deposition obviously corresponds to small standard deviations in the deposit distribution while the heterogeneous deposition corresponds to large standard deviations. Based on these features, a statistical model of pressure drop increase with deposition is developed, which is discussed next.

### 3. Statistical Model
Considering the random nature of the many factors affecting particle deposition and the large number density of pores, the statistical method is a natural choice to tackle the present problem.

At a given time moment during filtration and a given layer of pores of thickness $L$, we assume

(1). An open pore will be clogged when its specific deposit $\sigma$, that is the volume of deposits per unit volume of filter medium, exceeds a critical value $\sigma_c$.

(2). The specific deposit $\sigma$ of the open pores satisfy a truncated Gaussian distribution with nominal average $\bar{\sigma}$ and standard deviation $\lambda$ as Fig. 2 shows. The Gaussian distribution is truncated since for open pores $0 < \sigma < \sigma_c$. The probability density function of $\sigma$ is

$$\phi(\sigma; \bar{\sigma}, \lambda, \sigma_c) = \frac{1}{\sqrt{2\pi}\lambda I} e^{-\frac{(\sigma-\bar{\sigma})^2}{2\lambda^2}}, \quad (1)$$

where

$$I = \frac{1}{2}\left[erfc\left(-\frac{\bar{\sigma}}{\sqrt{2}\lambda}\right) - erfc\left(\frac{\sigma_c - \bar{\sigma}}{\sqrt{2}\lambda}\right)\right]. \quad (2)$$

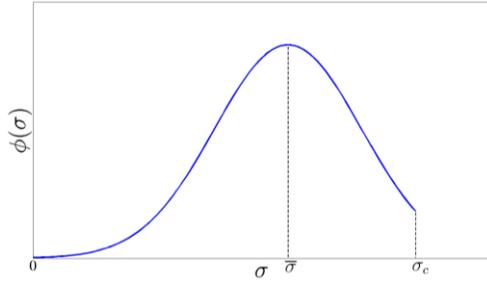

Fig. 2. The truncated Gaussian distribution

Of course, according to inference (3) of the two deposition regimes in Section 2, $\lambda$ is small for homogeneous deposition and large for heterogeneous deposition.

(3). Although $\bar{\sigma}$ keeps increasing during filtration, we assume $\lambda$ to be a constant through the filtration process.

(4). As the pressure gradient $\Delta p/L$ required to maintain a constant flow rate increases with time because more and more pores are clogged, the critical specific deposit $\sigma_c$ also increases, and the following formula is assumed to hold

$$\frac{\sigma_c}{\sigma_{c0}} = \frac{\Delta p/L}{(\Delta p/L)_0}, \tag{3}$$

where the subscript 0 denotes the corresponding values at the beginning of the filtration process when the filter medium is clean.

The reasoning underlying these assumptions is as follows. For assumption (1), since experiments [7] showed a distinct difference between open and clogged pores in terms of their contribution to pressure drop increase, it is expected that the transition from the open to clogged state occurs rather abruptly at a specific critical $\sigma$ value.

For assumption (2), the Gaussian distribution hypothesis is backed by the well-known central limit theorem [20] which indicates a physical quantity that is the sum of many statistically independent processes usually follows normal distribution closely. The amount of deposit is obviously such a physical quantity since it is the outcome of many statistically independent random variables like local fluid speed and particle concentration, grain size and shape, organism concentration in the medium etc.

Assumption (3) seems unreasonable at the first glance since for example in heterogeneous deposition we should expect the standard deviation of the deposition distribution to increase with time rather than being a constant. However, as we will see later the present model indeed is insensitive to the change in standard deviation as long as the filtration process remains in its original deposition regime.

As for assumption (4), physical intuition may convince one it is reasonable that as flow speeds up through open pores because of increased pressure gradient $\Delta p/L$, it is more difficult for them to clog, thus implying a higher critical specific deposit $\sigma_c$.

In a short time period, the mean specific deposit in open pores increases from $\bar{\sigma}$ to $\bar{\sigma} + d\bar{\sigma}$; as a consequence, the Gaussian distribution of the specific deposit in open pores translates to the right with the same displacement as shown in Fig. 3. The increase in the number fraction of clogged pores $m$ is then

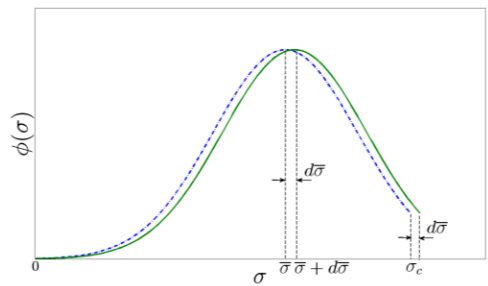

Fig. 3. Displacement of the truncated Gaussian distribution

$$dm = n\phi(\sigma_c; \bar{\sigma}, \lambda, \sigma_c)d\bar{\sigma}, \tag{4}$$

where $n = 1 - m$ is the number fraction of the open pores. The effective cross-sectional area of the layer decreases as the number fraction of open pores decreases and as a result the effective superficial velocity $u_s$, which is inversely proportional to the number fraction of open pores, increases. Using the Kozeny-Carman equation [8],

$$\frac{\Delta p}{L} = 180 \frac{u_s \mu}{d_p^2} \frac{(1-\epsilon)^2}{\epsilon^3}, \tag{5}$$

where $\mu$ is the fluid viscosity and $d_p$ is the diameter of particles, we conclude that the pressure gradient across this layer is also inversely proportional to the number fraction of open pores:

$$\frac{\Delta p/L}{(\Delta p/L)_0} = \frac{n_0}{n}, \tag{6}$$

and of course the initial number fraction of open pores $n_0 = 1$. Referring to Eqn. (3), we have

$$\frac{\sigma_c}{\sigma_{c0}} = \frac{n_0}{n}. \tag{7}$$

By taking the derivative of Eqn. (7), we have

$$\frac{d\sigma_c}{\sigma_{c0}} = -\frac{n_0}{n^2}dn = -\frac{\sigma_c}{\sigma_{c0}}\frac{dn}{n}, \tag{8}$$

That is,

$$d\sigma_c = \frac{\sigma_c}{n}(-dn). \tag{9}$$

Obviously

$$-dn = dm = n\phi(\sigma_c; \bar{\sigma}, \lambda, \sigma_c)d\bar{\sigma}, \tag{10}$$

whence

$$d\sigma_c = \sigma_c \phi(\sigma_c; \bar{\sigma}, \lambda, \sigma_c)d\bar{\sigma}. \tag{11}$$

Referring to inference (2) of the two-deposition-regime concept in Section 2, for the homogeneous deposition once a pore is clogged, it no longer actively collects particles, and its specific deposit value stays at the critical specific deposit value when it was clogged. Therefore in this regime the mean specific deposit of all pores at a particular layer is

$$\sigma_m = n\bar{\sigma} + \int_0^m \sigma_c dm. \tag{12}$$

From Eqn. (7) one may readily find that Eqn. (12) is equivalent to

$$\frac{\sigma_m}{\sigma_{c0}} = \frac{\bar{\sigma}}{\sigma_c} + \ln\left(\frac{\sigma_c}{\sigma_{c0}}\right). \tag{13}$$

In the heterogeneous deposition regime a clogged pore may have even higher particle collection ability than an open pore, therefore in this regime the distinction between open and clogged pores is only meaningful when pressure drop is concerned and is unnecessary if we just consider the deposit distribution. The specific deposits of open and clogged pores thus form one complete Gaussian distribution with mean $\bar{\sigma}$. Therefore in this regime

$$\sigma_m = \bar{\sigma}. \tag{14}$$

Equations (11), (13), (14) relate the mean specific deposit $\sigma_m$ with the critical specific deposit $\sigma_c$, which through the passage of Eqn. (3), is related to the pressure gradient $\Delta p/L$.

Before solving these equations, it is desirable to first render them in non-dimensional form. Let

$$\hat{\sigma}_c = \frac{\sigma_c}{\sigma_{c0}}; \quad \hat{\bar{\sigma}} = \frac{\bar{\sigma}}{\sigma_{c0}}; \quad \hat{\sigma}_m = \frac{\sigma_m}{\sigma_{c0}}; \quad \hat{\lambda} = \frac{\lambda}{\sigma_{c0}}, \tag{15}$$

Equations (11), (13), (14) then become

$$\frac{d\hat{\sigma}_c}{d\hat{\bar{\sigma}}} = \hat{\sigma}_c \phi(\hat{\sigma}_c; \hat{\bar{\sigma}}, \hat{\lambda}, \hat{\sigma}_c) = \frac{\hat{\sigma}_c}{\sqrt{2\pi}\hat{\lambda}I} e^{-\frac{(\hat{\sigma}_c - \hat{\bar{\sigma}})^2}{2\hat{\lambda}^2}}, \tag{16}$$

where

$$I = \frac{1}{2}\left[erfc\left(-\frac{\hat{\bar{\sigma}}}{\sqrt{2}\hat{\lambda}}\right) - erfc\left(\frac{\hat{\sigma}_c - \hat{\bar{\sigma}}}{\sqrt{2}\hat{\lambda}}\right)\right]. \tag{17}$$

For homogeneous deposition (as $\hat{\lambda}$ is small),

$$\hat{\sigma}_m = \frac{\hat{\bar{\sigma}}}{\hat{\sigma}_c} + \ln(\hat{\sigma}_c). \tag{18}$$

For heterogeneous deposition (as $\hat{\lambda}$ is large),

$$\hat{\sigma}_m = \hat{\bar{\sigma}}. \tag{19}$$

The initial conditions are

As $\hat{\bar{\sigma}}_m = 0, \hat{\bar{\sigma}} = 0$ and $\hat{\sigma}_c = 1$. \tag{20}

Equations (16) through (20) impose a non-linear initial-value problem and can be solved by using the Runge-Kutta method. The numerical solutions are presented and discussed in the next section.

## 4. Solutions and Discussions

Figure 4 shows solutions of the present model at four different $\hat{\lambda}$ values, in which Eqn. (18) is used for the two small $\hat{\lambda}$ values ($\hat{\lambda} = 0.01$ and $\hat{\lambda} = 0.1$) and Eqn. (19) is used for the two large $\hat{\lambda}$ values ($\hat{\lambda} = 10$ and $\hat{\lambda} = 100$).

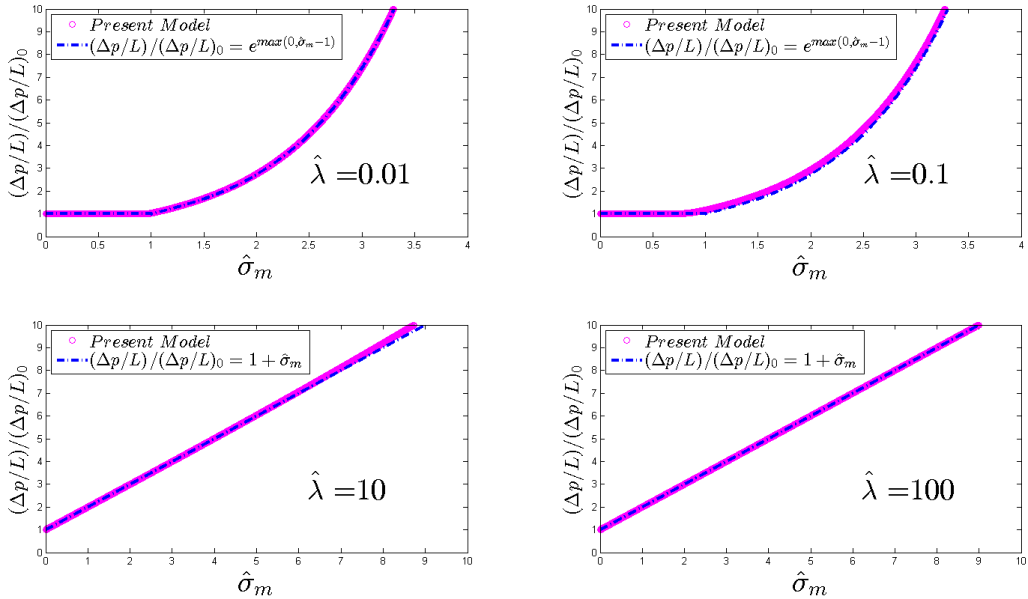

Fig. 4. Numerical solutions of the present model

One may immediately recognize the surprising features of these solutions. Firstly, although Eqns. (16) through (20) are highly non-linear, the solutions are linear for large

$\hat{\lambda}$ values; secondly, the same group of equations generates solutions with totally different behaviors: as $\hat{\lambda}$ is small (homogeneous deposition), the pressure drop does not change until the mean specific deposit $\sigma_m$ exceeds the initial critical specific deposit $\sigma_{c0}$, afterwards the pressure drop rises exponentially with specific deposit; on the contrary, as $\hat{\lambda}$ is large (heterogeneous deposition), the pressure drop increases linearly with specific deposit from the very beginning of filtration process. These characteristics are in excellent agreement with the two observed pressure increase patterns in granular filters [1][2][3][4][5][6][7]. Thirdly, the solutions of the present model are independent of $\hat{\lambda}$ in each deposition regime, which justifies proposition (3) of the present model (see Section 3), which assumes a constant $\hat{\lambda}$ during the whole filtration process.

The solutions of the present model can be summarized in the following formulae:
For homogeneous deposition (as $\hat{\lambda}$ is small):

$$\frac{\Delta p/L}{(\Delta p/L)_0} = \begin{cases} 1 & \text{if } \sigma_m < \sigma_{c0}; \\ \exp\left(\frac{\sigma_m}{\sigma_{c0}} - 1\right) & \text{if } \sigma_m \geq \sigma_{c0}. \end{cases} \qquad (21)$$

For heterogeneous deposition (as $\hat{\lambda}$ is large):

$$\frac{\Delta p/L}{(\Delta p/L)_0} = 1 + \frac{\sigma_m}{\sigma_{c0}}. \qquad (22)$$

A few numerical experiments show that the range of validity of Eqn. (21) is $\hat{\lambda} \leq 0.1$ and that of Eqn. (22) is $\hat{\lambda} \geq 1$. The model behavior in different $\hat{\lambda}$ ranges is discussed in Appendix B.

Although Eqns. (21) and (22) only give pressure drop increase of one single layer of pores, they can be easily extended to predict the total pressure drop increase across the whole granular filter.

For homogeneous deposition the pressure drop and deposition are concentrated to a thin stratum in the granular bed. As layers of pores in this stratum begin to clog, the pressure drop begins to rise. And these clogging layers should be rather similar, although not exactly the same, in extent of deposition. The reason is, according to Eqns. (6) and (21) the number of open pores of a layer decreases exponentially with specific deposit when this layer begins to clog. If there exists not-very-small difference in deposition level between two successive clogging layers, unless the particle concentration drops more quickly than exponentially from the heavier deposited layer to the lighter deposited layer, which is unlikely since the former now has much less active particle collecting pores than the latter, the latter will collect more particles than the former, and the deposition level difference between them will decrease and become very small. Therefore, since all these clogging layers in the thin stratum are similar, the mean pressure gradient and mean deposition level of the whole stratum are close to those of a single layer in it and Eqn. (21) is valid for the total pressure drop across the whole stratum and in turn, the whole filter since the pressure drop and deposition are concentrated to this stratum:

$$\frac{\Delta p}{(\Delta p)_0} = \begin{cases} 1 & if\ \sigma_m < \sigma_{c0}; \\ \exp\left(\frac{\sigma_m}{\sigma_{c0}} - 1\right) & if\ \sigma_m \geq \sigma_{c0}. \end{cases} \qquad (23)$$

$\Delta p$ and $\Delta p_0$ now denote the total pressure drops and $\sigma_m$ and $\sigma_{c0}$ are the bulk mean specific deposit and initial bulk critical specific deposit over the whole filter bed.

For heterogeneous deposition, the total pressure drop over the whole filter bed is

$$\sum_{i=1}^{N} \Delta p_i = \left(\frac{\Delta p}{L}\right)_0 L \sum_{i=1}^{N}\left(1 + \frac{\sigma_m}{\sigma_{c0}}\right)_i = \left(\frac{\Delta p}{L}\right)_0 (NL)\left[1 + \frac{\sum_{i=1}^{N}(\sigma_m)_i}{N\sigma_{c0}}\right], \qquad (24)$$

where subscript $i$ denotes the $i^{th}$ layer and $N$ is the total number of layers over the whole filter bed. Since $N\Delta p_0$ is the initial total pressure drop and $\frac{\sum_{i=1}^{N}(\sigma_m)_i}{N}$ is the bulk mean specific deposit, we have

$$\frac{\Delta p}{\Delta p_0} = 1 + \frac{\sigma_m}{\sigma_{c0}}, \qquad (25)$$

where again $\Delta p$ and $\Delta p_0$ denote the total pressure drops and $\sigma_m$ and $\sigma_{c0}$ are the bulk mean specific deposit and initial bulk critical specific deposit over the whole filter bed.

The quantitative accuracy of the present model is verified by comparing it with experimental data.

Figure 5 compares the homogeneous deposit solution of the present model, Eqn. (23),

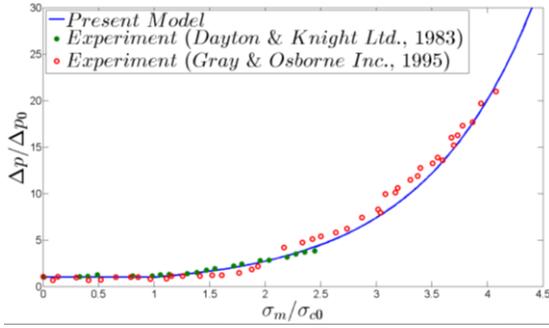

Fig. 5. Comparison of the present model with experimental data of slow sand filters

with two sets of experimental data. One data set was taken on a pilot slow sand water filter built in the Village of 100 Mile House, British Columbia of Canada by Dayton & Knight Ltd. [1] in 1983. Another set of data were obtained on a pilot slow sand filter built in City of Roslyn, Washington of U.S.A. by Gray & Osborne Inc. [2] in 1995. The present model matches both experimental data sets nicely.

Figure 6 shows a comparison of the present model, Eqn. (25) with the experimental data taken on a granular aerosol filter [5], which used a packed bed of pea gravels to collect char particles laden in air stream. In this experiment the bulk critical specific deposit was determined by inspecting the history of filter collection efficiency and the value was $\sigma_{c0} \approx 0.017$ [5]. The amount of deposit was obtained by sifting collected char particles from the pea gravels. The measured pressure drop values are very close to the theoretical prediction of Eqn. (25).

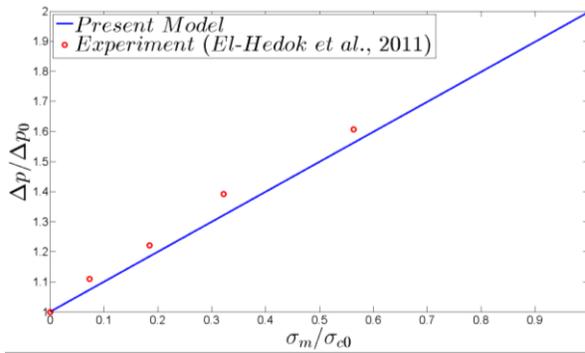

Fig. 6. Comparison of the present model with experimental data of a granular aerosol filter

5. Conclusions

The current study presented a consistent theory to explain the observed pressure drop increase with deposition behavior in granular filters. Two new concepts, homogeneous and heterogeneous depositions, were proposed, which could qualitatively interpret the drastic difference in pressure increase characteristics of different filtration processes. Based on these new concepts, a statistical model was developed by assuming truncated Gaussian distribution of pore deposit distribution and pressure drop being proportional to critical specific deposit. These assumptions resulted in a non-linear differential equation. Solution to this non-linear equation, however, is linear as the standard deviation of deposit distribution is large and is constant then exponential if the standard deviation of deposit distribution is small. These solutions agreed perfectly with experimental observations. Therefore the apparently very different pressure drop increase patterns in different granular filtration processes could be explained by one theory. The very unique mathematical behavior of the present statistical model may also be of interest to researchers working in wider areas.

The final conclusion is: the dramatically different pressure drop increase patterns in different granular filtration processes are not due to different pressure drop increase mechanisms, which in fact are the same (mainly clogging), but because of different deposit distributions among pores in the granular medium, and, of course, the physical/biological mechanisms that lead to such differences.

**Acknowledgements**

Z. Q. would like to thank Dr. Luiza Campos of University College London for the very intriguing discussions exchanged during the preparation of this work.

## Appendix A
### A possible mechanism that may lead to the two deposition regimes

As is well known [8], in a granular filter the single collector efficiency $\eta_s$, which measures the particle capture ability of a single unit in the granular medium (e.g. a pore), usually increases with flow speed as the flow speed is low because of enhanced particle inertial impaction, then decreases with flow speed as the flow speed becomes high due to growing possibility of particle bouncing-off from granule surfaces. The variation of $\eta_s$ with flow speed is sketched in Fig. 7.

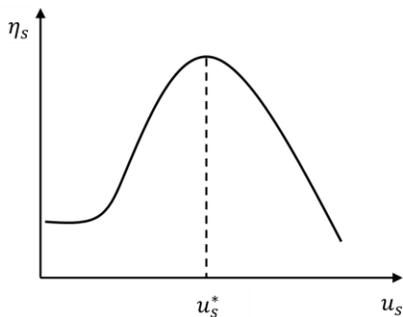

Fig. 7. Single collector efficiency variation with flow speed

Since heavier deposition results in higher flow resistance and lower flow speed, the collector efficiency of a more deposited pore becomes smaller than that of a less deposited pore if the flow speeds of pores are less than the critical speed $u_s^*$, which causes a homogeneous deposition distribution among the pores; if the flow speeds of pores are higher than $u_s^*$, the collector efficiency of a heavier deposited pore becomes even higher than that of a lighter deposited pore and this results in a heterogeneous deposition distribution among pores. From this brief analysis one may see filters with low flow speed, for instance slow sand filters, likely operate in the homogeneous deposition regime and filters with high flow speed, granular aerosol filters for example, may present heterogeneous deposition characteristics.

## Appendix B

## Model behavior in various $\hat{\lambda}$ ranges

Figure 8 shows how the homogeneous deposition solution evolves as $\hat{\lambda}$ increases. The

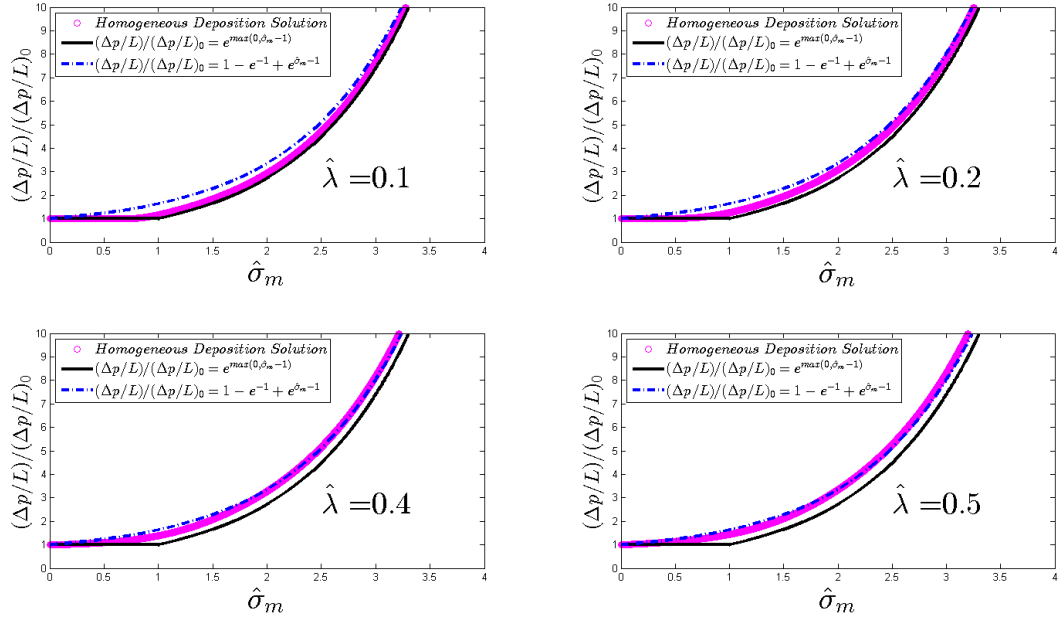

Fig. 8. Homogeneous deposition solutions

solutions shift from Eqn. (21) to the following form as $\hat{\lambda}$ increases from 0.1 to 0.5:

$$\frac{\Delta p/L}{(\Delta p/L)_0} = 1 - \exp(-1) + \exp\left(\frac{\sigma_m}{\sigma_{c0}} - 1\right). \tag{26}$$

The solutions are again independent of $\hat{\lambda}$ as $\hat{\lambda} \geq 0.5$.

Figure 9 shows the heterogeneous deposition solutions at various $\hat{\lambda}$ values. As $\hat{\lambda}$ drops to about 1, Eqn. (22) is still roughly held; however as $\hat{\lambda}$ continues to decrease, the solutions transfer to the following equation and this transition is completed at $\hat{\lambda} = 0.1$.

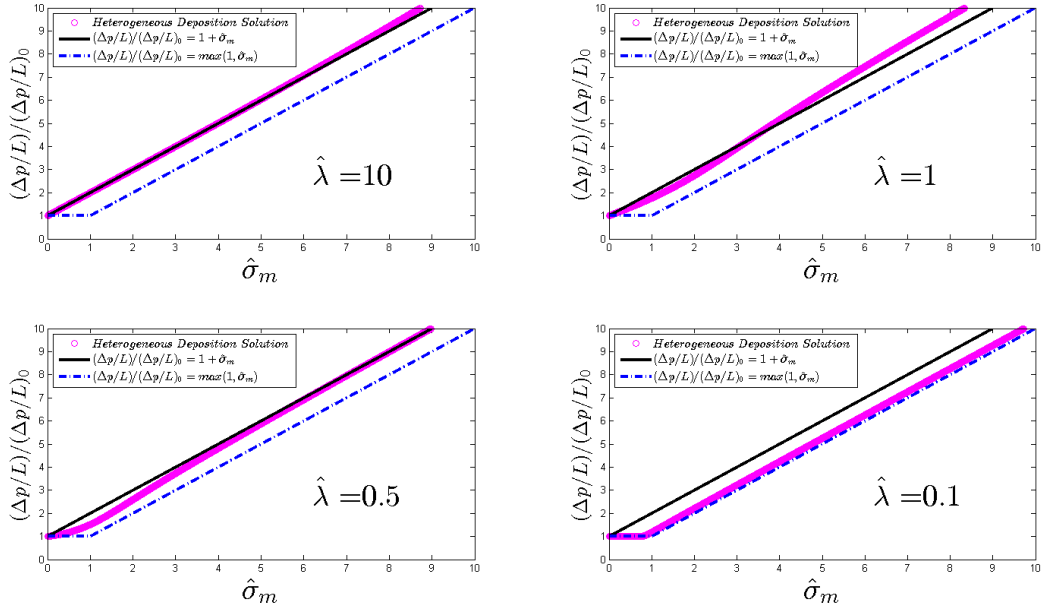

Fig. 9. Heterogeneous deposition solutions

$$\frac{\Delta p/L}{(\Delta p/L)_0} = \begin{cases} 1 & \text{if } \sigma_m < \sigma_{c0}; \\ \frac{\sigma_m}{\sigma_{c0}} & \text{if } \sigma_m \geq \sigma_{c0}. \end{cases} \qquad (27)$$

Again, the solutions are independent of $\hat{\lambda}$ as $\hat{\lambda} \leq 0.1$.

Based on physical reasoning it is very unlikely to observe homogeneous depositions as $\hat{\lambda} \gg 1$ nor heterogeneous depositions as $\hat{\lambda} \ll 1$. However, it is not clear if such solutions can take place in reality if $\hat{\lambda}$ is around 1.